\documentstyle[LTpaper,epsf,rotate]{article}
\newcommand{\be}{\begin{equation}}
\newcommand{\ee}{\end{equation}}

\newcommand{\et}{{\it et al.\ }}

\begin{document}
\title{\vspace*{-50pt}
\centerline{\sc XXI Int.\ Conf.\ on Low Temperature Physics in Prague, 
8-14 August 1996}\\[28pt]
Unconventional Pairing in Heavy Fermion Metals
\thanks{
We thank the Max Planck Gesellschaft and the Alexander von Humboldt
Stiftung for support. JAS also acknowledges partial support from the
NSF through the Science and Technology Center for Superconductivity
(DMR 91-20000).
}
}
\author{J. A. Sauls$^a$ and D. Rainer$^b$}

\address{\mbox{$^a$ Department of Physics \& Astronomy, Northwestern
University, Evanston, IL 60208, USA.}\\
\mbox{$^b$ Physikalisches Institut, Universit\"at Bayreuth,
D-95440 Bayreuth, Germany.}}

\abstract{
The Fermi-liquid theory of superconductivity is applicable to a broad
range of systems that are candidates for unconventional pairing, {\it
e.g.} heavy fermion, organic and cuprate superconductors.
Ginzburg-Landau theory provides a link between the thermodynamic
properties of these superconductors and Fermi-liquid theory. The
multiple superconducting phases of UPt$_3$ illustrate the role that is
played by the Ginzburg-Landau theory in interpreting these
novel superconductors. Fundamental differences between unconventional
and conventional anisotropic superconductors are illustrated by the
unique effects that impurities have on the low-temperature transport
properties of unconventional superconductors.  For special classes of
unconventional superconductors the low-temperature transport
coefficients are {\it universal}, i.e. independent of the impurity
concentration and scattering phase shift. The existence of a universal
limit depends on the symmetry of the order parameter and is achieved at
low temperatures $k_B T \ll \gamma \ll \Delta_0$, where $\gamma$ is the
bandwidth of the impurity induced Andreev bound states.  In the case of
UPt$_3$ thermal conductivity measurements favor an $E_{1g}$ or $E_{2u}$
ground state. Measurements at ultra-low temperatures should 
distinguish different pairing states.
}
\maketitle\pagestyle{empty}

\section{Introduction}

Theoretical investigation of {\it unconventional} pairing began with
the publication by  Anderson and Morel \cite{and61}, ``Generalized
Bardeen-Cooper-Schrieffer States and the Proposed Low-Temperature Phase
of $^3$He''. They studied the physical consequences of BCS pairing with
non-zero angular momentum, including superfluid phases with
spontaneously broken time-reversal symmetry. When superfluidity was
discovered \cite{osh72} in $^3$He it was immediately clear that this
was not a conventional s-wave BCS superfluid because there was more
than one superfluid phase. Further evidence for unconventional pairing
came from many experimental results and shortly after their discovery
the three superfluid phases of $^3$He were undisputedly identified as
p-wave spin-triplet superfluids.

The search for {\it unconventional superconductivity}, the  metallic
analog of superfluidity in $^3$He, was given a boost by the discoveries
of superconductivity in the class of heavy-fermion metals \cite{ste79},
particularly in the U-based compounds of UBe$_{13}$, UPt$_3$,
URu$_2$Si$_2$, UNi$_2$Al$_3$, and UPd$_2$Al$_3$.  Unusual temperature
dependences of the heat capacity, penetration depth, and sound
absorption led to conjectures that these materials were unconventional
superconductors \cite{ott87}. Much more experimental information is now
available, and there is consensus that some
heavy-fermion superconductors (if not all of them) show unconventional
pairing.  Interest in unconventional superconductivity expanded further
with reports of several experiments on cuprate superconductors that
supported earlier predictions \cite{bic87} of ``d-wave pairing''.

A rigorous classification of superconductors by the angular momentum of
the Cooper pairs (i.e. s-wave, p-wave, d-wave pairing, etc.) is not
appropriate in crystalline materials because angular momentum is not a
good quantum number. However, the terms ``d-wave pairing'', etc. are
often used interchangeably with ``unconventional pairing'' for states
in  which the pairing amplitude spontaneously breaks one or more
symmetries of the crystalline phase, i.e.  ${\cal R}*\Delta({{\cal
R}*\bf p}_f)\ne \Delta({\bf p}_f)$, where ${\cal R}\in {\cal G}= {\cal
G}_{spin}\times{\cal G}_{space}\times{\cal T} \times{\cal U}_{gauge}$
represents an operation of the full symmetry group ${\cal G}$ other
than a pure gauge transformation, and
\be\label{pair-amp}
\Delta_{\alpha\beta}({\bf p}_f)\sim
\left<a_{{\bf p}_f\alpha}a_{-{\bf p}_f\beta}\right>
\,,
\ee
is the equal-time pair amplitude, or order parameter, that describes
BCS-type superconductors.  Note that ${\bf p}_f$ is the momentum of a
quasiparticle on the Fermi surface and $\alpha,\beta$ are the spin
labels of the paired quasiparticles.  Fermion statistics requires that
the pair amplitude obey the anti-symmetry condition,
$\Delta_{\alpha\beta}({\bf p}_f)=-\Delta_{\beta\alpha}(-{\bf p}_f)$.

Essentially all of the candidates for unconventional superconductivity,
including the heavy-fermion and cuprate superconductors, have inversion
symmetry. This has an important consequence; the pairing interaction
that drives the superconducting transition decomposes into even- and
odd-parity sectors \cite{and84,vol84}. Thus, $\Delta_{\alpha\beta}({\bf
p}_f)$ necessarily has even or odd parity unless there is a second
superconducting instability into a state with different parity.

\medskip
\noindent{\it The d-wave model for the cuprates}
\medskip

Superconductivity in the high T$_c$ cuprates is generally believed to
result from pairing correlations within the CuO planes. Knight shift
measurements below T$_c$ indicate that the pairs form spin singlets
\cite{tak89}, and therefore even-parity orbital states. Unlike the
superfluid phases of $^3$He the cuprates exhibit a single
superconducting phase. The absence of multiple superconducting phases
in the cuprates suggests that the orbital pairing state belongs to one
of the four one-dimensional representations illustrated in Fig. 1, for
tetragonal crystal symmetry.

\bigskip
\begin{minipage}{\hsize}
\epsfxsize=\hsize
\hspace{-5mm}\epsfbox{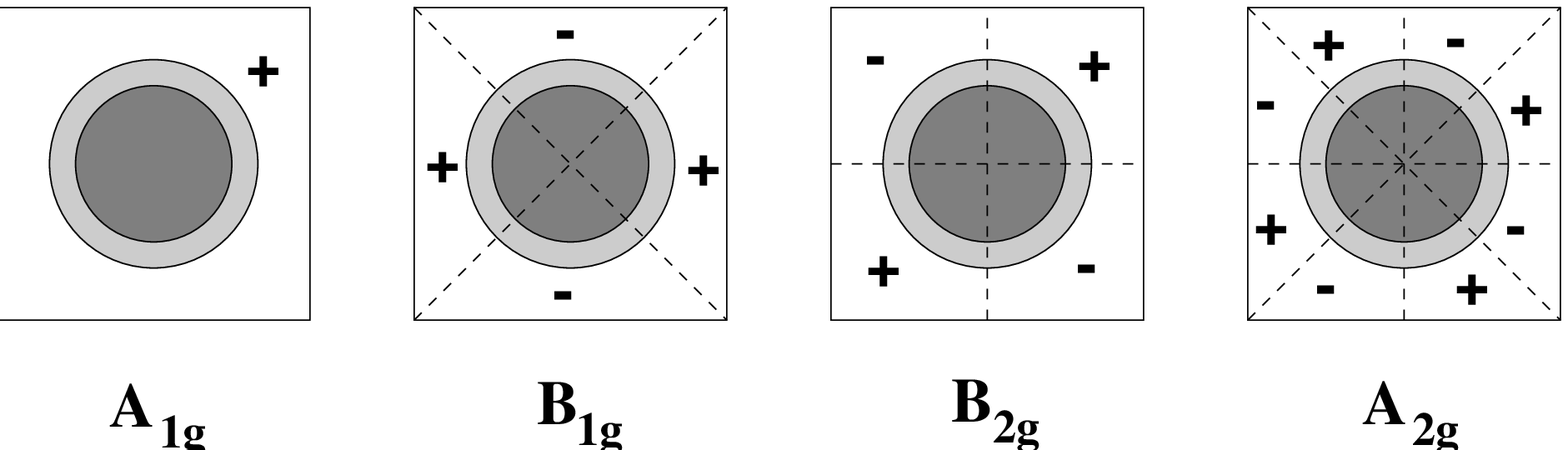}
\end{minipage}

\hspace*{-0.04\hsize}
\begin{minipage}{0.92\hsize}\small
Fig. 1.\hspace{2mm} Graphical representation the 1-dimensional basis
functions for even-parity tetragonal superconductors. The broken
reflection planes and the signs of the pairing state are indicated.
\end{minipage}
\medskip

The popular $d_{x^2-y^2}$ model for the cuprate superconductors
\cite{sca96} belongs to the $B_{1g}$ representation. This state breaks
the reflection symmetries for the (110) planes. As a result the order
parameter changes sign in momentum space as indicated in Fig. 1, and
the excitation gap $|\Delta({\bf p}_f)|$ has nodes in the $(110)$
direction on the Fermi surface. Both features lead to novel
transport properties in the superconducting state at very low
temperatures \cite{gra95,gra96}.

\medskip
\noindent{\it Multi-component Models of UPt$_3$}
\medskip

Considerable evidence in support of an unconventional pairing state in
the heavy-fermion materials has accumulated from specific heat, upper
critical field and various transport measurements, all of which show
anomalous properties compared to those of conventional superconductors
(see Refs. (\cite{ott87,hef96}) for original references).
However, some of the strongest evidence for unconventional
superconductivity comes from the multiple superconducting phases of
UPt$_3$ \cite{bru90,ade90}, which are a strong indication of a
multi-component pairing amplitude that necessarily breaks one or more
symmetries of the normal state.

There are several important features of the H-T phase diagram. (i)
There are two zero-field superconducting phases with a difference in
transition temperatures of ${\Delta T_c}/{T_c}\simeq 0.1$. (ii) A
change in slope of the upper critical field (a `kink' in
$H_{c2}^{\perp}$) is observed for ${\bf H}\perp\hat{\bf c}$. (iii)
There are three flux phases, and the phase transition lines separating
the flux phases appear to meet at a tetracritical point for all
orientations of ${\bf H}$ relative to $\hat{\bf c}$.  Three basic
models that have been proposed to explain the phases of UPt$_3$.

\noindent 1. {\it Multi-component Order Parameter coupled to a
Symmetry-Breaking Field (SBF)}. These are models based on a primary
order parameter belonging to a higher dimensional representation of the
symmetry group of the normal state.  For UPt$_3$, which has hexagonal
($D_{6h}$) symmetry, there are four 2D representations, $E_{1g(u)}$ and
$E_{2g(u)}$, with basis functions that transform like, ${\cal
Y}_{E_{1}\pm}\sim p_z(p_x\pm i p_y)$ for the even-parity $E_{1g}$
representation, and ${\cal Y}_{E_{2}\pm}\sim p_z(p_x\pm i p_y)^2$ for
the odd-parity $E_{2u}$ states.  The order parameter is then a complex
two-component vector $\vec{\eta}=(\eta_1,\eta_2)$ that transforms
according to the relevant 2D representation, and is related to the
pairing amplitude by
\begin{equation}
\Delta({\bf p}_f)=\eta_{+}{\cal Y}_{E_{1,2}+}
+
\eta_{-}{\cal Y}_{E_{1,2}-}
\,,
\end{equation}
with $\eta_{\pm}=\eta_1 \pm i\eta_2$. Multiple superconducting phases
correspond to different stationary solutions of the free energy
functional for $\vec{\eta}$.

The small splitting of the double transition in UPt$_3$ ($\Delta
T_c/T_c\simeq 0.1$) suggests the presence of a small symmetry breaking
energy scale and an associated lifting of a degeneracy of the possible
superconducting states belonging to the 2D representation
\cite{joy88,hes89,mac89}. The second zero-field transition just below
$T_c$ in UPt$_3$, as well as the anomalies observed in the upper and
lower critical fields, have been explained in terms of a weak symmetry
breaking field (SBF) that lowers the crystal symmetry from hexagonal to
orthorhombic, and consequently reduces the 2D E$_2$ (or E$_1$)
representation to two 1D representations with slightly different
transition temperatures. The key point is that right at $T_c$ all
states belonging to the 2D representation are degenerate, thus any SBF
that couples to $\vec{\eta}$ in second-order, and prefers a particular
state, will dominate very near $T_c$. At lower temperatures the SBF
energy scale, $\Delta T_c$, is a small perturbation compared to the
fourth-order terms in the fully developed superconducting state and one
recovers the results of the GL theory for the 2D representation with
small perturbations to the order parameter.

The phase diagram determined by ultrasound velocity measurements
indicates that the phase boundary lines meet at a tetracritical point
for both ${\bf H}||\hat{\bf c}$ and ${\bf H}\perp\hat{\bf c}$. This has
been argued to contradict the GL theory based on a 2D order parameter
\cite{luk91,mac91,che93}. The difficulty arises from gradient terms of
the form, $\left[(D_x\eta_1)(D_y\eta_2)^* + (D_x\eta_2)(D_y\eta_1)^*
+c.c.\right]$, that couple the two components of the order parameter.
These terms lead to `level repulsion' effects in the linearized GL
differential equations which prevent the crossing of two $H_{c2}(T)$
curves, corresponding to different superconducting phases. This feature
of the 2D model has spawned alternative models, designed specifically
to eliminate the `level repulsion' effect \cite{mac91,che93}, and
to a more specific version of the 2D model coupled to a SBF
based on the $E_{2u}$ representation \cite{sau94}. 

\noindent 2. {\it Accidental degeneracy of two order parameters}. These
models are based on {\it two} primary order parameters that are
unrelated by symmetry, i.e. belong to different irreducible
representations of the symmetry group, and are accidentally nearly
degenerate. The model of two 1D representations, $A_2$ and $B_1$ (``AB
model'') is a specific example \cite{che93}. The motivation behind this
model is that by choosing the two representations appropriately one can
guarantee that the `level repulsion' terms in the Ginzburg-Landau
equations for $H_{c2}$ are absent by symmetry \cite{luk91,che93}. What
is required is that the two order parameters corresponding to the two
irreducible representations, have different signatures under
reflection, or parity.  In this case the second-order gradient coupling
between the two order parameters vanishes, and the apparent
tetracritical point is present for all field orientations. The drawback
is that the accidental degeneracy models provide no explanation for the
near degeneracy of $T_c$, the observed correlation between $\Delta T_c$
and the AFM order parameter, or the pressure-temperature phase
diagram.

\noindent 3. {\it Enlarged Symmetry Group Models}.  These are {\it
hybrid} models that assume an accidental degeneracy in the form of a
larger symmetry group for the normal phase than one expects based on
the atomic and electronic structure of the crystal. A SBF is then
invoked to lift the degeneracy of a higher dimensional representation
for the larger symmetry group. By assuming a larger group than $D_{6h}$
one can again eliminate the `level repulsion' terms exactly in the GL
limit by judicious choice of the primary representation. Two different
versions of the enlarged symmetry group model have been proposed; one
based on an enlarged orbital symmetry group and the other based on the
assumption of {\it no} spin-orbit coupling. The orbital model starts
from the full rotation group, ${\cal G}_{space}=SO(3)$ and produces
multiple phases by crystal field splitting of the pairing states
\cite{zhi93}. This is a flexible model, but there is no strong
evidence to support treating UPt$_3$ as an isotropic material, even
approximately; and there is no explanation for the correlation between
the AFM order and the multiple superconducting phases.

The spin-channel version of this model \cite{mac91} assumes a symmetry
group composed of independent orbital and spin rotations with ${\cal
G}_{spin}=SU(2)$. The multi-component order parameter in this model
corresponds to the three spin-triplet amplitudes defined in terms of a
3D complex $\vec{\bf d}$ vector. The `level repulsion' terms are absent
for a 1D orbital representation. Coupling of the AFM order parameter to
the $\vec{\bf d}$ vector is proposed to split the transitions for the
different triplet sub-states. This model requires weak spin-orbit
coupling, which is at odds with theoretical estimates for the
spin-orbit coupling energy \cite{hef96}. Furthermore, weak spin-orbit
coupling is in contradiction with the observation of anisotropic Pauli
limiting \cite{shi86b}; a spin-triplet order parameter with weak
spin-orbit coupling would not exhibit Pauli limiting for any field
orientation (see also Ref.\cite{sau94}).

\medskip
\noindent{\it More on the SBF model for UPt$_3$}
\medskip

In the absence of accidental near degeneracy, a SBF is essential for
lifting the degeneracy of the pairing states near $T_c$ and producing
multiple superconducting phases \cite{hes89,mac89,joy90}. A natural
candidate for a SBF in UPt$_3$ is the AFM order in the basal plane
\cite{aep88}. In this case the GL functional includes a coupling of the
AFM order parameter to the superconducting order parameter; ${\cal
F_{{\rm SBF}}}\left[{\vec{\;\eta}}\right]
=\varepsilon\,M_s^2\,\int\limits d^{3}x\,\left(|\eta_1|^2 -
|\eta_2|^2\right)$, where $M_s$ is the AFM order parameter and the
coupling parameter $\varepsilon M_s^2$ determines the magnitude of the
splitting of the superconducting transition. The analysis of this GL
theory, including the SBF, is given in Ref. (\cite{hes89}); some of the
main results summarized below.

A double transition is predicted with a  splitting of $T_c$
proportional to $\Delta T_c\propto M_s^2$.  Support for the SBF model
of the double transition comes from pressure studies of the
superconducting and AFM phase transitions.  Heat capacity measurements
by Trappmann, {\et}\cite{tra91} show that both zero-field transitions
are suppressed under hydrostatic pressure, and that the double
transition disappears at $p_{*}\simeq 4\,kbar$.  Neutron scattering
experiments reported by Hayden, {\et}\cite{hay92} show that AFM order
disappears on the same pressure scale, at $p_c\simeq 3.2\,kbar$.

The low temperature phase ($T<T_{c*}$) is predicted to have broken
${\cal T}$ symmetry. This phase is doubly degenerate:
$\vec{\eta}_{\pm}\sim (a(T),\pm i\,b(T))$, reflecting the two
orientations of the internal orbital momentum of the ground state.

\smallskip
\begin{minipage}{\hsize}
\epsfysize=0.8\hsize
\epsfxsize=0.92\hsize
\hspace{-5mm}{\epsfbox{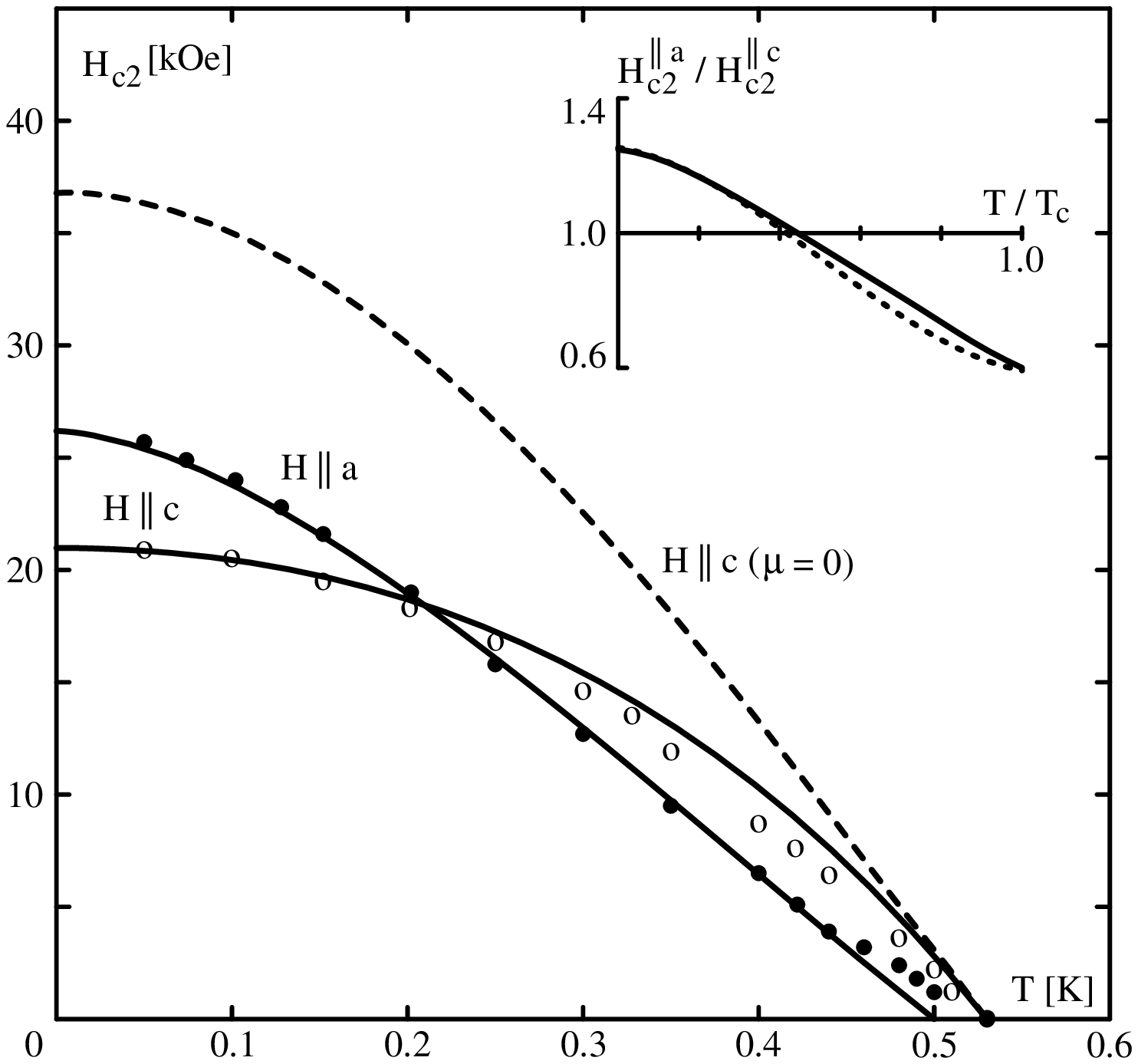}}
\end{minipage}
\smallskip

\hspace*{-0.04\hsize}
\begin{minipage}{0.92\hsize}\small
Fig. 2.\hspace{5mm} Calculations of $H_{c2}^{\perp}$ and
$H_{c2}^{\parallel}$ vs. $T$ for an odd-parity, triplet pairing state
with the $\vec{\bf d}$ vector locked to the $\hat{\bf c}$ direction
\cite{cho91}. The inset shows the ratio $H_{c2}^{\perp} /
H_{c2}^{\parallel}$ vs. $T/T_c$.  Note that $H_{c2}^{||}$ is suppressed
by paramagnetism ($\mu\approx\mu_B$) at low temperatures, while
$H_{c2}^{\perp}$ is independent of the paramagnetic coupling. The data
of Shivaram, et al. \cite{shi86b} are shown as the open and closed
circles.
\end{minipage}
\medskip

Another key feature of the UPt$_3$ phase diagram is the unusual
anisotropy of $H_{c2}$ shown in Fig. 2. The low-temperature anisotropy
is explained in terms of anisotropic Pauli limiting of an odd-parity,
spin-triplet state with the $\vec{\bf d}$-vector parallel to the
$\hat{\bf c}$ direction \cite{cho91}. This restricts one to the
E$_{2u}$ or E$_{1u}$ symmetry classes among the four possible 2D
representations.

The SBF is essential for producing an apparent tetracritical point, and
at a semi-quantitative level, can account for the magnitudes of the
slopes near the tetracritical point \cite{sau94,par96}.  However, there
are open questions regarding both the nature of the AFM order and its
coupling to superconductivity \cite{lus96c}.  As this model for UPt$_3$
illustrates, GL theory provides a central link between experiments, in
this case the phase diagram, and the more microscopic Fermi-liquid
theory which we discuss below.

\section{Fermi-Liquid Theory of Heavy Fermion\newline Superconductivity}

Conduction electrons in metals interact strongly with each other and
with the lattice. These interactions lead to correlations among the
electrons, and we have to view conduction electrons, in general, as a
system of correlated fermions. A particularly important system of
strongly correlated electrons are the conduction electrons of heavy
fermion metals. There is strong evidence that these heavy electrons
form a Landau Fermi liquid at low temperatures, and that their
superconducting states are well described by the Fermi liquid theory of
superconductivity. In the following we give a brief introduction to
this theory, then discuss recent applications to transport in heavy
fermion superconductors with unconventional pairing.

Landau showed that an ensemble of strongly interacting fermions may be
described by a distribution function for quasiparticle excitations, and
that this distribution function obeys a classical transport equation,
the Boltzmann-Landau transport equation.  It took more than 10 years
after Landau's theory of normal Fermi liquids, and the breakthrough in
the theory of superconductivity by BCS, to establish a complete
Fermi-liquid theory of superconductivity.  Earlier general theories of
superconductivity lacked the quasiclassical aspects of Landau's
Fermi-liquid theory. The first complete quasiclassical (QC) theory of
superconductivity was formulated in a series of publications by
Eilenberger \cite{eil68}, Larkin and Ovchinnikov \cite{lar69} and
Eliashberg \cite{eli72}. It is presented and discussed in several
review articles \cite{ser83,lar86,rai95}. The quasiclassical
theory allows one to calculate all superconducting phenomena of
interest, including transition temperatures, excitation spectra,
Josephson effects, vortex structures, the electromagnetic response,
etc. In this theory the dynamical degrees of freedom of electronic
quasiparticles are described partly by classical mechanics,
and partly by quantum statistics. The classical degrees of freedom are
the motion in ${\bf p}$-${\bf R}$ phase space; i.e.  quasiparticles
move along classical trajectories. Quantum degrees of freedom are the
spin of a quasiparticle and the particle-hole degree of freedom, which
form a four-dimensional Hilbert space of ``internal degrees of
freedom''.

\medskip
\begin{minipage}{\hsize}
\epsfxsize=0.9\hsize
\hspace{10mm}\epsfbox{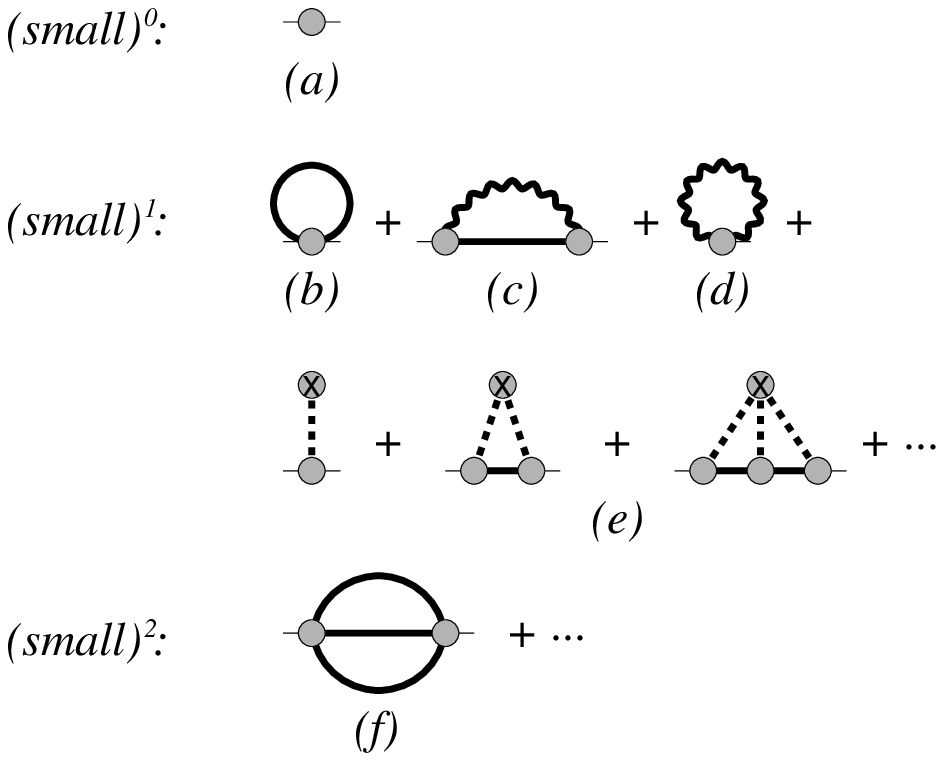}
\end{minipage}\smallskip

\hspace*{-0.04\hsize}
\begin{minipage}{0.92\hsize}\small
Fig. 3.\hspace{5mm} Leading order self-energy diagrams of Fermi-liquid
theory. The vertices (shaded circles) represent the sum of all
high-energy processes and give rise to interactions between the
quasiparticles (smooth propagator lines), phonons (wiggly propagator
lines) and impurities (dashed lines). The order in the parameter
'small' $\sim\epsilon/E_f$ is indicated for each diagram.
\end{minipage}
\medskip

Derivations of the Boltzmann-Landau transport equation from first
principles \cite {lan59,eli62} use many-body Green's function
techniques, and lead to explicit expressions for the various terms of
the transport equation in terms of self-energies. The self-energies
describe the effects of electron-electron, electron-phonon and
electron-impurity scattering. The complete set of diagrams for the
leading order self-energies are shown in Fig. 3; the filled circles are
vertices representing the effective lattice potential, which determines
the quasiparticle Fermi surface and Fermi velocities, as well as
quasiparticle-quasiparticle, quasiparticle-phonon, and
quasiparticle-impurity interactions. We follow Landau and consider
these vertices as phenomenological parameters of the Fermi-liquid
model, but in principle they can be obtained from the full many-body
theory. The selection  of leading order self-energies holds for the
normal and the superconducting states.  The vertices are not affected
in leading order by the superconducting transition; superconductivity
affects only the electron propagators. The special achievement of
Eilenberger, Larkin, Ovchinnikov, and Eliashberg was to convert Dyson's
equations for the electron propagators into transport equations for
quasiclassical propagators. The resulting equations define the
quasiclassical theory of superconductivity.

The central equation of the quasiclassical theory is the  
transport equation for the quasiclassical Keldysh propagator, $\hat
g^{K}$,
\begin{eqnarray}\label{qcltr} 
&\left(\epsilon\hat\tau_3-\hat v-\hat
\sigma^R\right)\otimes
\hat g^K-
\hat g^K\otimes
\left(\epsilon\hat\tau_3-\hat v-\hat
\sigma^A\right)- \nonumber\\
&\hat\sigma^K\otimes
\hat g^A+ \hat g^R
\otimes 
\hat\sigma^K+i\hbar{\bf v}_f\cdot\mbox{\boldmath{$\nabla$}}
\hat g^K=0\,.
\end{eqnarray}
The retarded, advanced, and Keldysh propagators, $\hat g^{\,R,A,K}({\bf
p}_f,{\bf R};\epsilon,t)$, as well as the self-energies,
$\hat\sigma^{\,R,A,K}({\bf p}_f,{\bf R};\epsilon,t)$, and the external
potentials, $\hat{v}({\bf p}_f,{\bf R},t)$, are 4$\times$4 Nambu
matrices, acting on the 4D Hilbert space of internal degrees of
freedom. The $\otimes$-product stands for the usual $4\times 4$-matrix
product and a product in the energy-time variables. Details of this
compact notation are explained in the reviews of the quasiclassical
theory \cite{ser83,lar86,rai95}. The determination of $\hat g^K$
from the transport equation requires knowledge of the external
potentials, the advanced, retarded and Keldysh self-energies, and the
advanced and retarded quasiclassical propagators.  These propagators
are auxiliary quantities which in general have no direct physical
interpretation, except in the adiabatic limit \cite{ser83} where they
determine the local quasiparticle density of states. The retarded and
advanced propagators are solutions of
\begin{equation}\label{qcltrRA} 
\left[\epsilon\hat\tau_3-\hat v-\hat
\sigma^{R,A},
\hat g^{R,A}\right]_{\otimes} 
+i\hbar{\bf v}_f\cdot\mbox{\boldmath{$\nabla$}} \hat g^{R,A}=0\ .
\end{equation}
The physically relevant set
of solutions of Eqs. (\ref{qcltr}) and (\ref{qcltrRA})
must satisfy the normalization conditions,
\begin{equation}\label{norm1} 
\hat g^{R,A}\otimes 
\hat g^{R,A}=-\pi^2\hat 1\ , \ \ 
\hat g^{R}\otimes 
\hat g^{K}+ 
\hat g^{K}\otimes 
\hat g^{A}=0\ .
\end{equation} 

Measurable quantities such as the charge current density, ${\bf j}({\bf
R},t)$, can be calculated from the diagonal components of the
quasiclassical propagator, $\hat g^K$. For example, the charge current
density is given by
\begin{equation}\label{ncur}
{\bf j}({\bf R},t)= 
\int{2\,d^2{\bf p}_f\over (2\pi)^3\mid {\bf v}_f\mid}
\int{d\epsilon\over 4\pi i}\,
e{\bf v}_f \ g^K({\bf p}_f,{\bf R};\epsilon,t)
\,.
\end{equation}
The quasiclassical equations (\ref{qcltr})-(\ref{norm1}) are
supplemented by self-consistency equations for the the quasiclassical
self-energies. These equations are shown in diagrammatic notation in
Fig. 3. They include, for example, the ``gap equation'', which is the
self-consistency equation for the off-diagonal self-energy. Explicit
forms of the quasiclassical self-energies can be found in Refs.
\cite{ser83,lar86,rai95}.

The quasiclassical theory is especially well suited for studying
unconventional superconductors.  Some of the results of the theory for
conventional s-wave superconductors are simply generalized to
unconventional superconductors.  For example, the density of states
$N({\bf p}_f; \epsilon)$ of a homogeneous spin-singlet superconductor
in equilibrium has the standard BCS form, $N({\bf p}_f; \epsilon)=
N_f\Re[\epsilon/\sqrt{\mid\Delta\mid^2-\epsilon^2}]$, but with the
isotropic gap $|\Delta|$ replaced by an anisotropic gap, $|\Delta({\bf
p}_f)|$, for each point on the Fermi surface.

However, inhomogeneous and non-equilibrium situations exhibit more
striking differences between conventional and unconventional
superconductors that reflect both the broken symmetries of the order
parameter and the coherence properties of the superconducting state.

\medskip
\noindent{\it Particle-Hole Coherence in QC Theory}
\medskip

Of special importance for understanding superconducting phenomena are
the quantum-mechanical ``internal degrees of freedom'', the {\sl spin}
and the {\sl particle-hole} degrees of freedom of an electron. Quantum
coherence between particle excitations and hole excitations is a key
feature of the BCS theory of superconductivity and the origin of all
non-classical effects in superconductors (e.g. supercurrents, coherence
factors in transition amplitudes, Andreev reflection, etc.).
Particle-hole coherence is incorporated into the quasiclassical theory
by grouping particle excitations (occupied one-electron states above
the Fermi energy) and hole excitations (empty one-electron states
below the Fermi energy) into a doublet. For clean superconductors, and
for long wavelength spatial variations, $\xi_0=\hbar v_f/2\pi
T_c\gg\hbar/p_f$, we can make a quasiclassical envelope approximation
for the Bogoliubov amplitudes; the resulting particle and hole
amplitudes obey Andreev's equation,\cite{and64}
\be
\left(\epsilon\hat{\tau}_3 - \hat{\Delta}({\bf p}_f, {\bf R})\right)
\,\vec{\varphi}_{{\bf p}_f}
+i\hbar{\bf v}_f\cdot\mbox{\boldmath{$\nabla$}}
\,\vec{\varphi}_{{\bf p}_f}=0
\,,
\ee
where $\vec{\varphi}_{{\bf p}_f}= \left( u_{{\bf p}_f} , v_{{\bf p}_f}
\right)$ are the quasiclassical particle ($u$) and hole ($v$)
amplitudes. Andreev's equation is a first-order differential equation
for excitations propagating along straight-line trajectories determined
by the Fermi velocity ${\bf v}_f({\bf p}_f)$ at a point ${\bf p}_f$ on
the Fermi surface. In the superconducting state, the order parameter
$\hat{\Delta}({\bf p}_f, {\bf R})$ mixes the normal-state excitations
coherently into two branches of particle-like (${\bf p}_f\cdot{\bf
v}_{+}>0$) and hole-like (${\bf p}_f\cdot{\bf v}_{-}<0$) excitations.

Coherence between particle and hole excitations leads to dramatic
effects on the excitation spectrum of an unconventional superconductor
in the vicinity of an impurity or a surface.
Consider an excitation that is incident on a specular surface from
the bulk region of a superconductor with broken reflection symmetry
perpendicular to the plane of the interface. When an excitation
reflects off the surface elastically its momentum shifts to a new point
on the Fermi surface, ${\bf p}_f\rightarrow\underline{\bf p}_f$. Thus,
the incident and reflected wavepacket propagate through different order
parameter fields, $\Delta({\bf p}_f,{\bf R})$ vs.
$\Delta(\underline{\bf p}_f,{\bf R})$. As a result, surface scattering
generally leads to Andreev 
scattering, a process of ``retro-reflection'' in which a
particle-like excitation undergoes branch conversion into a hole-like
excitation with reversed group velocity. Bound states may occur at
energies for which the phases of multiply-reflected particle- and
hole-like excitations interfere constructively.

The effects of a surface on particle-hole coherence is most pronounced
if the scattering induces a change in sign of the order parameter along
the classical trajectory. This occurs for a $d_{x^2-y^2}$
superconductor with a $(110)$ surface, an $E_{1g}$ or $E_{2u}$ order
parameter and a surface normal to the $\hat{\bf c}$ axis. If a sign
change of the order parameter occurs along the trajectory, then a
zero-energy bound state forms at the surface with equal amplitudes for
the particle and hole components \cite{buc81}.  These states are
expected to give rise to zero-bias anomalies in the conductance for NIS
tunnel junctions \cite{hu94,tan95,buc95b}.

\medskip
\noindent{\it Impurity-induced Andreev bound states}
\medskip

The novel effects of impurities in unconventional superconductors,
including pairbreaking, can be understood in similar terms. Fermion
bound states are formed by impurity-induced Andreev scattering. Figure
4 illustrates the connection between potential scattering and the
development of an Andreev bound state for s-wave impurities in the
$d_{x^2-y^2}$ model; similar physical processes apply to unconventional
models of the heavy fermion superconductors.  Impurities give rise to
elastic scattering of states near the Fermi surface, with transition
matrix elements given by $u({\bf p}_f, {\bf p}_f')$. The scattering
amplitude is determined by the ${t}$-matrix,
\noindent\begin{eqnarray}
\hat t^{R,A}({\bf p}_{\!f},{\bf p}_{\!f}^{\ \prime},{\bf R};\epsilon,t)
        = u({\bf p}_{\!f},{\bf p}_{\!f}^{\ \prime})
{+}N_{\!f}\!\!\int d{\bf k}_{\!f}
\nonumber
\\
u({\bf p}_{\!f},{\bf k}_{\!f})
\hat g^{R,A}({\bf k}_{\!f},{\bf R};\epsilon,t)\otimes
\hat t^{R,A}({\bf k}_{\!f},{\bf p}_{\!f}^{\ \prime},
        {\bf R};\epsilon,t)
\end{eqnarray}
The ${t}$-matrix sums two types of repeated scattering processes to all
orders: (i) potential scattering processes with a change in momentum,
but no change in the internal state, and (ii) Andreev scattering
processes, i.e. branch conversion with no change in momentum, but
reversal of the group velocity. The Andreev processes are {\it induced}
by the sign changes of the order parameter that result from scattering
around the Fermi surface as shown in Fig. 4. Bound states also form
from impurity-induced Andreev scattering, with an energy that depends
on the scattering phase shift. Neglecting broadening effects from the
continuum states above the gap, the bound state energy is given by
$\epsilon_{bound}=\Delta_0\cos{\delta_0}$. However, the continuum
excitations are nearly gapless in the vicinity of the nodes of
$\Delta({\bf p}_f)$, so bound states at finite energy are broadened
into resonances. For strong scattering the bound state occurs near zero
energy and the resonance width is narrow; e.g. in the unitarity limit
the impurity bound states appear at $\epsilon=0$ and are sharp.

\smallskip
\begin{minipage}{\hsize}
\epsfxsize=0.92\hsize
\epsfbox{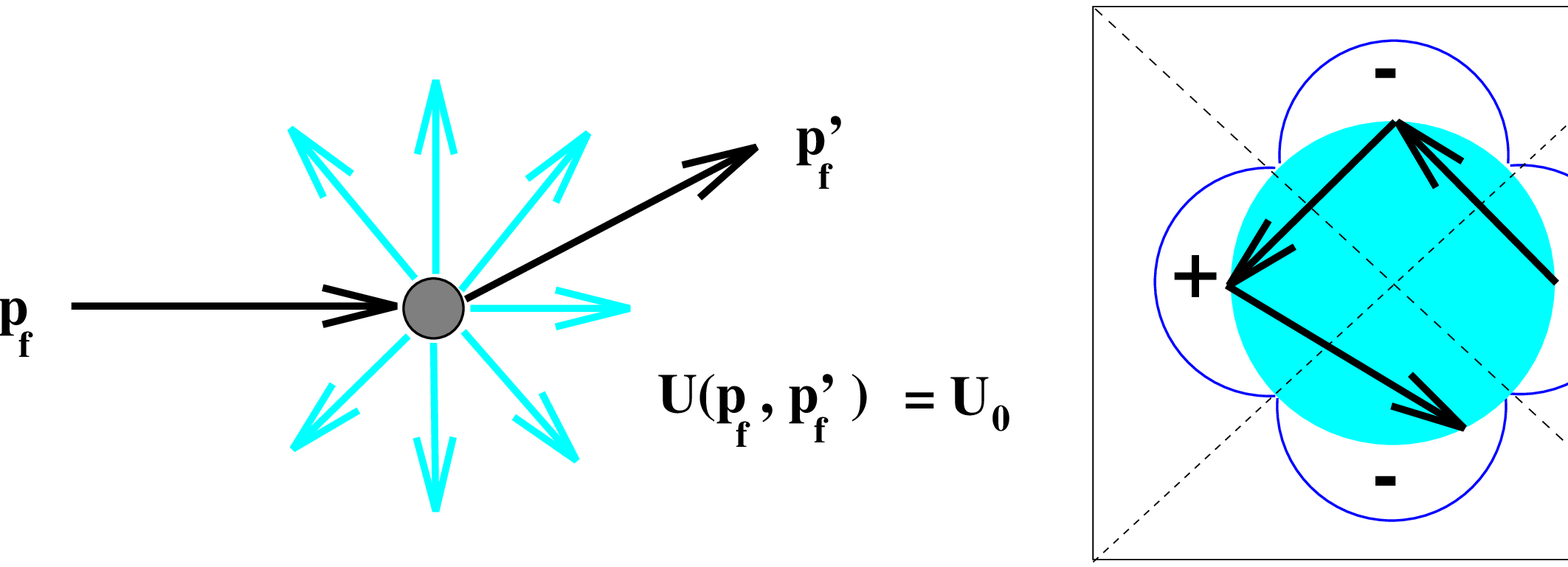}
\end{minipage}
\smallskip

\hspace*{-0.04\hsize}
\begin{minipage}{0.92\hsize}\small
Fig. 4.\hspace{5mm} Scattering of quasiparticles by an impurity induces
Andreev scattering, and for unconventional pairing states the formation
of Andreev bound states or resonances.
\end{minipage}
\medskip

In a dilute alloy a finite density of impurities leads to a finite
density of impurity states, an Andreev {\it band} \cite{cho87}. The
impurity bandwidth and density of states can be calculated from the
self-consistent $\hat{t}$-matrix and the leading order impurity
self-energy terms (diagrams $3e$). The bandwidth is given by
\begin{equation}
\gamma = \Gamma_u\,
{
\langle\gamma\left(|\Delta({\bf p}_{\!f})|^2
   + \gamma^2\right)^{-1/2}
\rangle
\over
\mbox{cot}^2\delta_0 + 
\langle\gamma\left(|\Delta({\bf p}_{\!f})|^2
     + \gamma^2\right)^{-1/2}
\rangle^2
}\,,
\end{equation}
where $\Gamma_u=n_{\rm imp}/\pi N_f$ is the maximum impurity scattering
rate in the the normal state for a fixed concentration, $n_{\rm imp}$,
of impurities and $\langle ...\rangle$ is an average over the Fermi
surface. In the unitarity limit the impurity bandwidth is
$\gamma\simeq \sqrt{\pi\Gamma_u/2\Delta_0}\sim\sqrt{n_{\rm imp}}$.

For $\gamma<\epsilon\ll\Delta_0$ the low energy spectrum is dominated
by continuum excitations in the vicinity of the nodal lines; for the
$E_{1g}$ and $E_{2u}$ models of UPt$_3$ $N(\epsilon)\approx
N_f(\epsilon/\Delta_0)$. Below $\epsilon\approx\gamma$ the impurity
band dominates with $N(\epsilon)\simeq N(0)\approx
N_f(\gamma/\Delta_0)$. These two contributions to the spectrum also
give rise to different features in the transport properties of the
heavy fermion superconductors \cite{gra95,gra96a} To calculate the
transport coefficients in the superconducting state we need the Keldysh
propagator for these branches of the excitation spectrum. Linear
response functions for electrical and thermal transport in
strong-coupling, dirty and unconventional superconductors are derived
from the quasiclassical transport equations in Refs.
\cite{rai95,gra95,gra96a}.

\section{Thermal Conductivity of UPt$_3$ - $E_{2u}$ Model}

The anisotropic thermal conductivity coefficients for the $E_{2u}$
model are shown in Fig. 5 as a function of temperature, and compared
with the data of Lussier, et al. \cite{lus96b} for $T\ge 0.1 T_c$.
The fits of the $E_{2u}$ model to the low temperature region,
$0.1<T/T_c<0.5$ are very good for both directions of heat flow.  More
detailed analysis shows that both the $E_{2u}$ and $E_{1g}$ pairing
states give excellent fits to the thermal conductivity for
$T>0.1\,T_c$, and that ultra-low temperature (i.e. $T<\gamma$) heat
transport measurements and impurity studies should distinguish these
pairing states \cite{gra96}.

\smallskip
\begin{minipage}{\hsize}
\epsfysize=0.8\hsize
\epsfxsize=0.92\hsize
\epsfbox{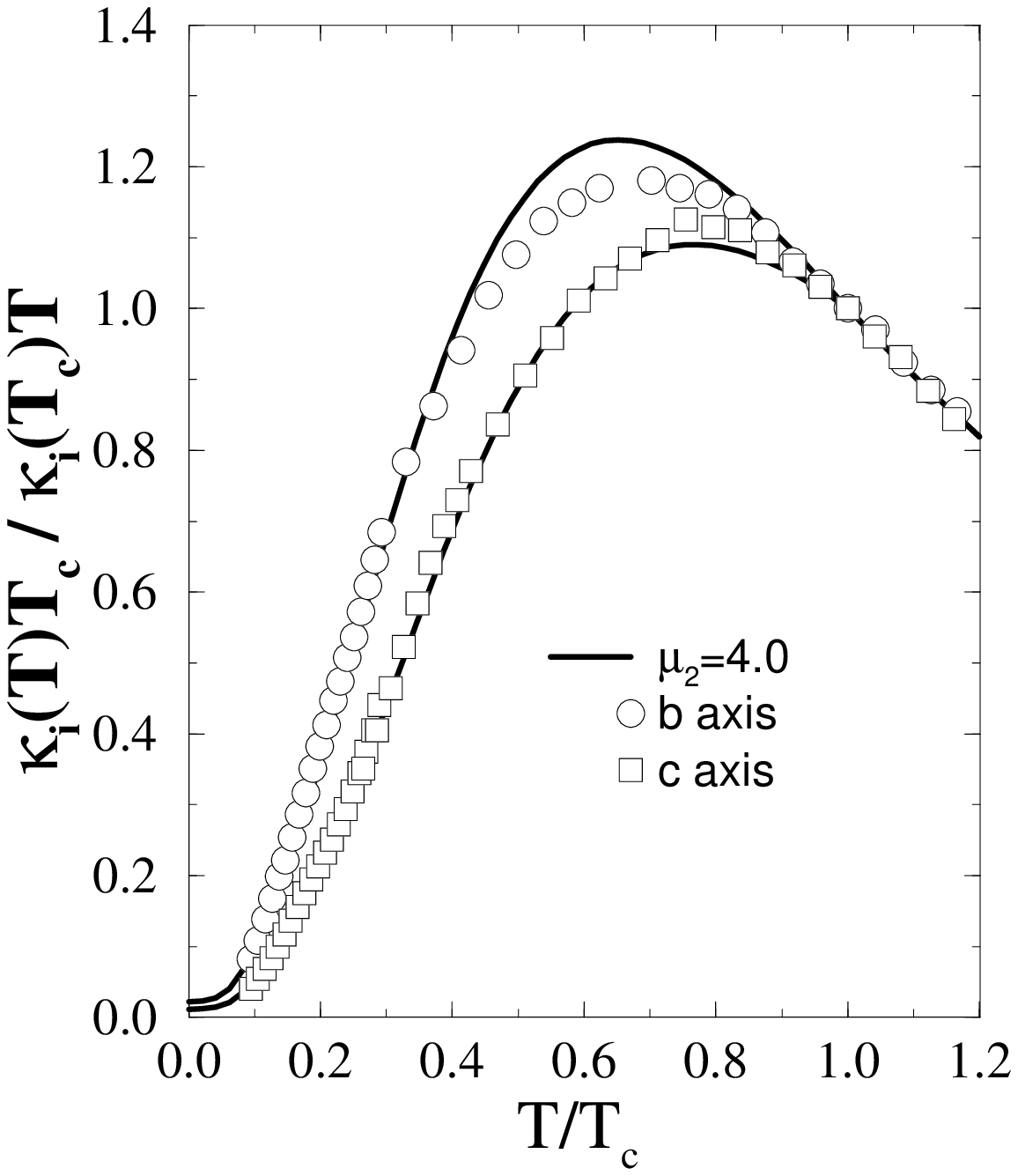}
\end{minipage}
\smallskip

\hspace*{-0.04\hsize}
\begin{minipage}{0.92\hsize}\small
Fig. 5.\hspace{5mm} The normalized thermal conductivity of the E$_{2u}$
ground state \cite{gra96} compared with the data of Ref. \cite{lus96b}.
The impurity scattering rate is $\Gamma_0=0.01\pi\,T_{c0}$ and the
phase shift is $\delta_0= 90^{\circ}$. The slope and curvature
parameters are $\mu_1=2.0$ and $\mu_2=4.0$.
\end{minipage}
\medskip

\noindent{\it Ultra-Low Temperature Heat Transport}

Signatures the pairing symmetry arise from both the low-energy
continuum states and the impurity-induced Andreev band. This novel
metallic band deep in the superconducting state gives rise to {\it
universal} transport coefficients for $k_B T\ll\gamma$ 
\cite{gra95}.
Whether or not a universal limit develops at low temperatures depends
sensitively on the nodal structure of the excitation gap. Both the
E$_{1g}$ and E$_{2u}$ models for the order parameter of UPt$_3$ lead to
an excitation gap with a nodal line in the basal plane and point nodes
along the $\pm\hat{\bf c}$ directions \cite{nor92}. The difference in
the excitation spectrum for these two states is that the gap opens
linearly for small angles away from the $\hat{\bf c}$ direction for the
E$_{1g}$ order parameter, but quadratically for E$_{2u}$ order
parameter; $\,|\Delta_{\rm E_{1g}}(\vartheta)| \simeq  \Delta_0\,
\mu_1|\vartheta|$ and $|\Delta_{\rm E_{2u}}(\vartheta)| \simeq
\Delta_0\, \mu_2|\vartheta|^2$. This difference is reflected in
$\lim_{T\rightarrow 0}\kappa_{c}/T$, which is universal for the
E$_{2u}$ model, but non-universal for the E$_{1g}$ model \cite{gra96},
\begin{eqnarray}
\kappa_{c}/T
\simeq\frac{\pi^2}{3}N_f^2v_f^2
\left(\frac{\gamma}{\mu_{1}^2\Delta_0^2}\right)
\left(1+\frac{a_{E_{1g}}^2T^2}{\gamma^2}\right)
\\
\kappa_{c}/T
\simeq\frac{\pi^2}{3}N_f^2v_f^2
\left(\frac{1}{2\mu_{2}\Delta_0}\right)
\left(1+\frac{a_{E_{2u}}^2T^2}{\gamma^2}\right)
\end{eqnarray}

For heat flow in the basal plane the thermal conductivity is determined
by the line node, and is universal for {\it both} $E_{1g}$ and $E_{2u}$
pairing states.  The leading temperature corrections to $\kappa/T$
are given by a Sommerfeld expansion for $T < \gamma < T_c$. The $T^3$
corrections are non-universal and in the unitarity limit,
$\gamma^2\propto n_{\rm imp}$, so the coefficient of the $T^3$ term
scales as $1/n_{\rm imp}$ for the E$_{2u}$ state and $1/\sqrt{n_{\rm
imp}}$ for the E$_{1g}$ state.
Observed  universality for $T\ll\gamma$ and the scaling of the $T^3$
correction with $n_{\rm imp}$ would be  strong tests of the
 symmetry of the  order parameter.

We thank M. J. Graf, D. W. Hess, and S.-K. Yip for  many
contributions to the work reported here.


\end{document}